\documentclass[preprint,superscriptaddress]{revtex4}
\usepackage{amsmath}
\usepackage{array}
\usepackage{graphicx}
\usepackage{textcomp}

\renewcommand{\tag}{\label}
\input{tcilatex}
\begin{document}

\title{{\Large Optical Soliton Propagation in a Free-Standing Nonlinear
Graphene Monolayer with Defects}}
\author{Frederick Ira Moxley III}
\affiliation{Hearne Institute for Theoretical Physics, Department of Physics \&
Astronomy, Louisiana State University, Baton Rouge, LA 70803, USA}
\author{Tim Byrnes}
\affiliation{National Institute of Informatics, 2-1-2 Hitotsubashi, Chiyoda-ku, Tokyo
101-8430, Japan}
\author{ Adarsh Radadia}
\affiliation{Institute for Micromanufacturing, Louisiana Tech University, Ruston, LA
71272, USA}
\author{Weizhong Dai}
\affiliation{Mathematics \& Statistics, College of Engineering \& Science, Louisiana Tech
University, Ruston, LA 71272, USA}

\begin{abstract}
Recently, optical soliton propagation in an intrinsic nonlinear graphene
monolayer configuration has been discovered. However, optical soliton
behavior in a free-standing graphene monolayer with defects has not yet been
studied. The objective of this article is to employ the generalized
finite-difference time-domain (G-FDTD) method to efficiently simulate bright
optical solitons, illustrating propagation of the electric field
distribution in a free-standing nonlinear layer with variation in
nonlinearity along its width. These variations of nonlinearity along the
width represent graphene impurities, or defects. Results show that solitons
propagate effectively even in the presence of strong spatial variations in
the nonlinearity, implying the robustness of the medium with respect to
optical propagation.
\end{abstract}

\maketitle

\section{Introduction}

Recently, the isolation of graphene monolayers from bulk graphite has
inspired many studies \cite{Novoselov}, including those related to optical
and opto-electronical applications \cite{Bonaccorso, Liu}. Findings show
that photon propagation can be better controlled in materials with
increasing nonlinear optical susceptibility \cite{Mikhailov, Ishikawa},
which allows formation of temporal and spatial electro-magnetic (EM)
solitons \cite{Kartashov, Liu07}. Such structures are of high interest in
optical communications. In the optical regime, Nesterov and co-workers \cite%
{Nesterov} have used finite element modeling (FEM) to demonstrate that 2D
graphene monolayers can support such a quantum phenomenon in the absence of
defects. However, more realistically the graphene will have inhomogeneties
in various parameters such as the interaction. Such parameter deviation can
cause optical scattering, leading to loss of the optical soliton for large
propagation distances. It is our understanding that such deviation in the
parameters is rather difficult to simulate using finite element methods as
they induce numerical instabilities, particularly for strong spatial
variations. The motivation of this study is to employ our recently developed
generalized finite-difference time-domain (G-FDTD) method \cite{Moxley13} to
solve a non-homogenous vector Helmholtz equation analogue. The G-FDTD scheme
was developed for solving linear and nonlinear Schr\"{o}dinger equations 
\cite{Moxley13,Moxley}. It is explicit, permits an accurate solution with
simple computation, and also satisfies a discrete analogue of the mass
conservation law for nonlinear Schr\"{o}dinger equations. In this study, we
will use the G-FDTD method to simulate graphene imperfections in a
free-standing configuration, permitting an analysis of the robustness of the
optical soliton propagation. We will investigate whether or not these
optical solitons propagate with great stability in the presence of very
large spatial inhomogeneties. This will illustrate the effectiveness of the
monolayer configuration for optical soliton propagation and allow us to
examine the robustness of the soliton propagation under non-ideal conditions.

\section{Helmholtz's Equation Approximation}

In this study, we consider a single graphene monolayer placed inside a
planar linear dielectric waveguide, as shown in Figure 1. The vertical
confinement along the $y-$axis provides a direction for the propagating EM
mode. We assume that graphene is a thin film of thickness $d_{gr}$ on the
order of nanometers. As seen in Figure 1, a three-dimensional free standing
graphene monolayer configuration will yield the same results described by
two-dimensional conductivity. Hence, the photon propagation of light inside
the nonlinear monolayer can be expressed as the non-homogeneous vector
Helmholtz's equation as follows \cite{Nesterov}: 
\begin{equation}
c^{2}\epsilon _{0}\left[ (\frac{n_{s}}{c})^{2}\frac{\partial ^{2}}{\partial
t^{2}}-\nabla ^{2}\right] \mathbf{A}(\mathbf{r},t)=\mathbf{j}_{NL}(\mathbf{r}%
,t),  \tag{1}
\end{equation}%
where $\mathbf{A}(\mathbf{r},t)$ is the magnetic potential vector, and $%
n_{s} $ is the linear refractive index of graphene. To solve Eq. (1), we
assume a solution of the form 
\begin{equation}
\mathbf{A}(\mathbf{r},t)=\frac{1}{2}\left\{ \mathbf{\hat{A}}(x)F(z,y)\exp
[i(\beta z-\omega t)]+c.c.\right\} ,  \tag{2}  \label{aequation}
\end{equation}%
where $\mathbf{\hat{A}}(x)$ is an arbitrary integrable function that governs
the confinement and polarization of the EM field along the $x$-direction,
and $c.c.$ is the complex conjugate of the former term. Moreover, the EM
field in the graphene plane is then governed by the complex function $F(z,y)$%
, and the propagation constant along the $z$-direction is given by $\beta $.
Now, we insert Eq. (2) into Eq. (1), apply the slowly varying amplitude
approximation, and project both sides of the resulting equation over the
transpose conjugate [$\mathbf{\hat{A}}(x)$]$^{\ast T}$. The auxiliary
function is taken to be $f(z,y)\equiv F(z,y)\exp (-i\phi z)$ (where $\phi
\equiv (k_{s}^{2}-\beta ^{2}+I_{2}/I_{1})/2$, $k_{s}=n_{s}^{2}\omega
^{2}/c^{2},$ $I_{1}\equiv \int\nolimits_{-\infty }^{\infty }|\mathbf{\hat{A}}%
(x)|^{2}dx$ and $I_{2}\equiv \int\nolimits_{-\infty }^{\infty }dx[\mathbf{%
\hat{A}}(x)]^{\ast T}\partial ^{2}\mathbf{\hat{A}}(x)/\partial x^{2})$. As
such, Eq. (1) can be solved in terms of the auxiliary function $f(z,y)$ as 
\begin{equation}
2i\beta \frac{\partial f(z,y)}{\partial z}+\frac{\partial ^{2}f(z,y)}{%
\partial y^{2}}+g\left\vert f(z,y)\right\vert ^{2}f(z,y)=0,  \tag{3}
\label{schrodinger}
\end{equation}%
where $g\equiv \frac{3}{4}\omega ^{4}\chi _{gr}^{(3)}I_{3}/I_{1}c^{2},$ and $%
I_{1}\equiv \int\nolimits_{-d_{gr}/2}^{d_{gr}/2}|\mathbf{\hat{A}}(x)|^{4}dx.$
Eq. (3) corresponds to the nonlinear Schr\"{o}dinger equation (NLSE), whose
bright soliton solutions have the form \cite{Boyd, Kivshar} 
\begin{equation}
f(z,y)=\frac{1}{w}\sqrt{\frac{2}{g}}\mbox{sech}(\frac{y}{w})\exp (\frac{iz}{%
2\beta w^{2}}).  \tag{4}  \label{exact}
\end{equation}%
Here, $w$ defines the soliton width and $\chi _{gr}^{(3)}$ is the Kerr-type
third-order effective nonlinear susceptibility. Solving the auxiliary
function this way permits a simple and effective way to solve Eq. (2), which
calculates the electric field as $\mathbf{E}(\mathbf{r},t)=-\partial \mathbf{%
A}(\mathbf{r},t)/\partial t$. The propagation of light in a waveguide by
which the refractive index contrasts between the core and the cladding is
described physically by Eqs. (3) and (4). More importantly, Eq. (4) shows
mathematically that optical solitons do indeed exist in graphene. From Eq.
(3), it can be seen that the interaction parameter $g$ governs the graphene
inhomogeneties. As such, a constant $g$ will represent an ideal graphene
monolayer configuration without defects. Similarly, a variable $g$ will
represent a more realistic graphene monolayer configuration that contains
impurities or defects. It is of interest to study such graphene impurities
to determine whether or not such deviations of the interaction parameter $g$
induce optical scattering of the propagating soliton, especially in the case
of large propagation distances. As pointed out in the previous section, for
a constant $g$, Nesterov and co-workers \cite{Nesterov} employed a finite
element modeling to demonstrate that 2D graphene monolayers can support
optical soliton propagation. In this study, we will investigate the optical
soliton propagation in the case of a variable $g$. Because such deviation in 
$g$ is rather difficult to simulate using finite element methods, we will
utilize the explicit G-FDTD method to solve Eq. (3) with a variable $g$.

\section{Explicit G-FDTD Scheme}

To apply the explicit G-FDTD method in \cite{Moxley, Moxley13} (the general
idea of the G-FDTD dates back to Visscher \cite{Visscher}), we first assume
that the auxiliary function $f(z,y)$ be a sufficiently smooth function which
vanishes for sufficiently large $\left\vert y\right\vert $, and split the
variable $f(z,y)$ into real and imaginary components, 
\begin{equation}
f(z,y)=f_{\text{real}}(z,y)+if_{\text{imag}}(z,y).  \tag{5}
\end{equation}%
Inserting Eq. (5) into Eq. (3) and then separating the real and imaginary
parts result in the following coupled set of equations: 
\begin{eqnarray}
\frac{\partial f_{\text{real}}(z,y)}{\partial z} &=&-\frac{1}{2\beta }%
\left\{ \frac{\partial ^{2}f_{\text{imag}}(z,y)}{\partial y^{2}}+g\left[ f_{%
\text{real}}^{2}(z,y)+f_{\text{imag}}^{2}(z,y)\right] f_{\text{imag}%
}(z,y)\right\}  \notag \\
&=&-\frac{1}{2\beta }\left( A+g\left\vert f\right\vert ^{2}\right) f_{\text{%
imag}}(z,y),  \TCItag{6}
\end{eqnarray}%
where $A=\frac{\partial ^{2}}{\partial y^{2}}$ and $\left\vert f\right\vert
^{2}=f_{\text{real}}^{2}(z,y)+f_{\text{imag}}^{2}(z,y)$.

We denote $f_{\text{real}}(k\Delta y,n\Delta z)$ as $f_{\text{real}}^{n}(k), 
$ and $f_{\text{imag}}(k\Delta y,n\Delta z)$ as $f_{\text{imag}}^{n}(k)$ for
simplicity. Using a Taylor series expansion at $z=(n-1/2)\Delta z,$ we
obtain 
\begin{equation}
f_{\text{real}}^{n}(k)-f_{\text{real}}^{n-1}(k)=2\sum\limits_{m=0}^{M}(\frac{%
\Delta z}{2})^{2m+1}\frac{1}{(2m+1)!}\frac{\partial ^{2m+1}f_{\text{real}%
}(y,z_{n-1/2})}{\partial z^{2m+1}}+O(\Delta z^{2M+3})  \tag{7a}
\end{equation}%
and 
\begin{equation}
f_{\text{imag}}^{n}(k)-f_{\text{imag}}^{n-1}(k)=2\sum\limits_{m=0}^{M}(\frac{%
\Delta z}{2})^{2m+1}\frac{1}{(2m+1)!}\frac{\partial ^{2m+1}f_{\text{imag}%
}(y,z_{n-1/2})}{\partial z^{2m+1}}+O(\Delta z^{2M+3}).  \tag{7b}
\end{equation}%
We then evaluate those derivatives in Eq. (7a) by using Eqs. (6a) and (6b)
repeatedly, where $z$ in $f_{\text{real}}^{2}(z,y)+f_{\text{imag}}^{2}(z,y)$
is fixed at $(n-1/2)\Delta z$: 
\begin{equation}
\frac{\partial f_{\text{real}}(y,z_{n-1/2})}{\partial z}=-\frac{1}{2\beta }%
\left( A+g\left\vert f^{n-1/2}\right\vert ^{2}\right) f_{\text{imag}}(y,z_{n-%
\frac{1}{2}}),  \tag{8a}
\end{equation}%
\begin{eqnarray}
\frac{\partial ^{3}f_{\text{real}}(y,z_{n-1/2})}{\partial z^{3}} &=&-\frac{1%
}{2\beta }\left( A+g\left\vert f^{n-1/2}\right\vert ^{2}\right) \frac{%
\partial ^{2}f_{\text{imag}}(y,z_{n-1/2})}{\partial z^{2}}  \notag \\
&=&\frac{1}{2\beta }\left( A+g\left\vert f^{n-1/2}\right\vert ^{2}\right) 
\frac{1}{2\beta }\left( A+g\left\vert f^{n-1/2}\right\vert ^{2}\right) \frac{%
\partial f_{\text{real}}(y,z_{n-1/2})}{\partial z}  \notag \\
&=&\left[ \frac{1}{2\beta }(A+g\left\vert f^{n-1/2}\right\vert ^{2})\right]
^{3}f_{\text{imag}}(y,z_{n-\frac{1}{2}}),  \TCItag{8b}
\end{eqnarray}%
\begin{eqnarray}
\frac{\partial ^{5}f_{\text{real}}(y,z_{n-1/2})}{\partial z^{5}} &=&\left[ 
\frac{1}{2\beta }(A+g\left\vert f^{n-1/2}\right\vert ^{2})\right] ^{3}\frac{%
\partial ^{2}f_{\text{imag}}(y,z_{n-1/2})}{\partial z^{2}}  \notag \\
&=&-\left[ \frac{1}{2\beta }(A+g\left\vert f^{n-1/2}\right\vert ^{2})\right]
^{3}\left[ \frac{1}{2\beta }(A+g\left\vert f^{n-1/2}\right\vert ^{2})\right] 
\frac{\partial f_{\text{real}}(y,z_{n-1/2})}{\partial z}  \notag \\
&=&-\left[ \frac{1}{2\beta }(A+g\left\vert f^{n-1/2}\right\vert ^{2})\right]
^{5}f_{\text{imag}}(y,z_{n-\frac{1}{2}}),  \TCItag{8c}
\end{eqnarray}%
and so on, where $\left\vert f^{n-1/2}\right\vert ^{2}=f_{\text{real}%
}^{2}(k\Delta y,z_{n-1/2})+f_{\text{imag}}^{2}(k\Delta y,z_{n-1/2})$.
Substituting Eq. (8) and other similar equations into Eq. (7a) gives 
\begin{eqnarray}
f_{\text{real}}^{n}(k)-f_{\text{real}}^{n-1}(k) &=&2\sum\limits_{m=0}^{M}(%
\frac{\Delta z}{2})^{2m+1}\frac{(-1)^{m}}{(2m+1)!}\left[ -\frac{1}{2\beta }%
(A+g\left\vert f^{n-1/2}\right\vert ^{2})\right] ^{2m+1}f_{\text{imag}%
}(y,z_{n-\frac{1}{2}})  \notag \\
&&+O(\Delta z^{2M+3}).  \TCItag{9}
\end{eqnarray}%
Similarly, using Eqs. (6a) and (6b) repeatedly to evaluate those derivatives
in Eq. (7b), we obtain 
\begin{equation}
\frac{\partial f_{\text{imag}}(y,z_{n-1/2})}{\partial z}=\frac{1}{2\beta }%
\left( A+g\left\vert f^{n-1/2}\right\vert ^{2}\right) f_{\text{real}}(y,z_{n-%
\frac{1}{2}}),  \tag{10a}
\end{equation}%
\begin{eqnarray}
\frac{\partial ^{3}f_{\text{imag}}(y,z_{n-1/2})}{\partial z^{3}} &=&\frac{1}{%
2\beta }\left( A+g\left\vert f^{n-1/2}\right\vert ^{2}\right) \frac{\partial
^{2}f_{\text{real}}(y,z_{n-1/2})}{\partial z^{2}}  \notag \\
&=&-\frac{1}{2\beta }\left( A+g\left\vert f^{n-1/2}\right\vert ^{2}\right) 
\frac{1}{2\beta }\left( A+g\left\vert f^{n-1/2}\right\vert ^{2}\right) \frac{%
\partial f_{\text{imag}}(y,z_{n-1/2})}{\partial z}  \notag \\
&=&-\left[ \frac{1}{2\beta }(A+g\left\vert f^{n-1/2}\right\vert ^{2})\right]
^{3}f_{\text{real}}(y,z_{n-\frac{1}{2}}),  \TCItag{10b}
\end{eqnarray}%
\begin{eqnarray}
\frac{\partial ^{5}f_{\text{imag}}(y,z_{n-1/2})}{\partial z^{5}} &=&-\left[ 
\frac{1}{2\beta }(A+g\left\vert f^{n-1/2}\right\vert ^{2})\right] ^{3}\frac{%
\partial ^{2}f_{\text{real}}(y,z_{n-1/2})}{\partial z^{2}}  \notag \\
&=&\left[ \frac{1}{2\beta }(A+g\left\vert f^{n-1/2}\right\vert ^{2})\right]
^{3}\left[ \frac{1}{2\beta }(A+g\left\vert f^{n-1/2}\right\vert ^{2})\right] 
\frac{\partial f_{\text{imag}}(y,z_{n-1/2})}{\partial z}  \notag \\
&=&\left[ \frac{1}{2\beta }(A+g\left\vert f^{n-1/2}\right\vert ^{2})\right]
^{5}f_{\text{real}}(y,z_{n-\frac{1}{2}}),  \TCItag{10c}
\end{eqnarray}%
and so on. Substituting Eq. (10) and other similar equations into Eq. (7b)
gives 
\begin{eqnarray}
f_{\text{imag}}^{n}(k)-f_{\text{imag}}^{n-1}(k) &=&2\sum\limits_{m=0}^{M}(%
\frac{\Delta z}{2})^{2m+1}\frac{(-1)^{m+1}}{(2m+1)!}\left[ -\frac{1}{2\beta }%
(A+g\left\vert f^{n-1/2}\right\vert ^{2})\right] ^{2m+1}f_{\text{real}%
}(y,z_{n-\frac{1}{2}})  \notag \\
&&+O(\Delta z^{2M+3}).  \TCItag{11}
\end{eqnarray}%
Noting that the term $\left\vert f^{n-1/2}\right\vert ^{2}$ in Eqs. (9) and
(11) needs to be evaluated, we use a similar argument and obtain 
\begin{eqnarray}
f_{\text{real}}^{n+1/2}(k)-f_{\text{real}}^{n-1/2}(k)
&=&2\sum\limits_{m=0}^{M}(\frac{\Delta z}{2})^{2m+1}\frac{(-1)^{m}}{(2m+1)!}%
\left[ -\frac{1}{2\beta }(A+g\left\vert f^{n}\right\vert ^{2})\right]
^{2m+1}f_{\text{imag}}(y,z_{n})  \notag \\
&&+O(\Delta z^{2M+3}),  \TCItag{12a}
\end{eqnarray}%
\begin{eqnarray}
f_{\text{imag}}^{n+1/2}(k)-f_{\text{imag}}^{n-1/2}(k)
&=&2\sum\limits_{m=0}^{M}(\frac{\Delta z}{2})^{2m+1}\frac{(-1)^{m+1}}{(2m+1)!%
}\left[ -\frac{1}{2\beta }(A+g\left\vert f^{n}\right\vert ^{2})\right]
^{2m+1}f_{\text{real}}(y,z_{n})  \notag \\
&&+O(\Delta z^{2M+3}),  \TCItag{12b}
\end{eqnarray}%
where $\left\vert f^{n}\right\vert ^{2}=f_{\text{real}}^{2}(k\Delta
y,z_{n})+f_{\text{imag}}^{2}(k\Delta y,z_{n})$. Next, we couple Eqs. (9),
(11) and (12) together, drop out the truncation error $O(\Delta z^{2M+3})$,
and replace $\frac{\partial ^{2}}{\partial x^{2}}$ by a fourth-order
accurate central difference operator, $\frac{1}{\Delta y^{2}}D_{y}^{2}u(k)=%
\frac{1}{12\Delta y^{2}}[-u(k+2)+16u(k+1)-30u(k)+16u(k-1)-u(k-2)]$. This
results in our explicit G-FDTD scheme for solving Eq. (3) as follows: 
\begin{equation}
f_{\text{real}}^{n}(k)-f_{\text{real}}^{n-1}(k)=2\sum\limits_{m=0}^{M}\frac{%
(-1)^{m}}{(2m+1)!}\left[ -\frac{1}{2\beta }(\frac{\sigma }{2}D_{y}^{2}+\frac{%
\lambda \Delta z}{2}\left\vert f^{n-1/2}\right\vert ^{2})\right] ^{2m+1}f_{%
\text{imag}}^{n-1/2}(k),  \tag{13a}
\end{equation}%
\begin{equation}
f_{\text{imag}}^{n}(k)-f_{\text{imag}}^{n-1}(k)=2\sum\limits_{m=0}^{M}\frac{%
(-1)^{m+1}}{(2m+1)!}\left[ -\frac{1}{2\beta }(\frac{\sigma }{2}D_{y}^{2}+%
\frac{\lambda \Delta z}{2}\left\vert f^{n-1/2}\right\vert ^{2})\right]
^{2m+1}f_{\text{real}}^{n-1/2}(k);  \tag{13b}
\end{equation}%
\begin{equation}
f_{\text{real}}^{n+1/2}(k)-f_{\text{real}}^{n-1/2}(k)=2\sum\limits_{m=0}^{M}%
\frac{(-1)^{m}}{(2m+1)!}\left[ -\frac{1}{2\beta }(\frac{\sigma }{2}D_{y}^{2}+%
\frac{\lambda \Delta z}{2}\left\vert f^{n}\right\vert ^{2})\right] ^{2m+1}f_{%
\text{imag}}^{n}(k),  \tag{13c}
\end{equation}%
\begin{equation}
f_{\text{imag}}^{n+1/2}(k)-f_{\text{imag}}^{n-1/2}(k)=2\sum\limits_{m=0}^{M}%
\frac{(-1)^{m+1}}{(2m+1)!}\left[ -\frac{1}{2\beta }(\frac{\sigma }{2}%
D_{y}^{2}+\frac{\lambda \Delta z}{2}\left\vert f^{n}\right\vert ^{2})\right]
^{2m+1}f_{\text{real}}^{n}(k);  \tag{13d}
\end{equation}%
where $\sigma =\Delta z/\Delta y^{2}$. Thus, once $f_{\text{imag}}^{n-1}(k)$%
, $f_{\text{real}}^{n-1}(k)$, $f_{\text{imag}}^{n-1/2}(k)$ and $f_{\text{real%
}}^{n-1/2}(k)$ are given, one may explicitly calculate $f_{\text{real}%
}^{n}(k)$ and $f_{\text{imag}}^{n}(k)$ using Eqs. (13a) and (13b), and then
obtain $f_{\text{real}}^{n+1/2}(k)$ and $f_{\text{imag}}^{n+1/2}(k)$ using
Eqs. (13c) and (13d). It is an explicit iteration, and therefore the
computation is simple and fast. The stability of the explicit G-FDTD scheme
for solving the NLSE and the discrete conservation law that the scheme
satisfies can be seen in \cite{Moxley13}. It can be seen that the truncation
error between the above scheme and Eqs. (9), (11) and (12) is $O(\Delta
y^{4}+\Delta z^{2M+3})$. However, the truncation error between the above
scheme and the original NLSE, Eq. (3), may be different because the $z$ in
the term $\left\vert f(z,y)\right\vert ^{2}$ is fixed in the small $z$
interval either $(z_{n-1},z_{n})$ or $(z_{n-1/2},z_{n+1/2})$ in the above
derivation. Furthermore, it should be pointed out that the partial
derivative $\frac{\partial ^{2}}{\partial y^{2}}$ can alternatively be
approximated using a spectral, or other higher-order method. In this study,
we confine our attention to the finite-difference method with a fourth-order
central difference approximation.

\section{Optical Soliton Simulation}

We first calculated the electric field $|\mathbf{E}(\mathbf{r},t)|$ along a
free-standing non-linear layer, where $g=5,$ $w=2,$ $\beta =-0.5,$ and $\phi
=\omega =1$ in Eqs. (1) - (4), where the parameters represent a
free-standing graphene monolayer configuration in the absence of defects.
This can be done by injecting optical solitons at one edge of the monolayer.
The initial condition was chosen to be a bright soliton based on the
analytical solution 
\begin{equation}
f(z,y)=\frac{1}{w}\sqrt{\frac{2}{g}}\mbox{sech}(\frac{y}{w})\exp (\frac{iz}{%
2\beta w^{2}}),  \tag{14}  \label{initconds}
\end{equation}%
and $\mathbf{\hat{A}}(x)=\frac{1}{1+x^{2}}$ in Eq. (\ref{aequation}). The
purpose of this calculation was for comparison with the results of Nesterov
et al. \cite{Nesterov}. In our computation, the domain was taken to be $%
-10\leq x,y\leq 10,$ $0\leq z\leq 1.$ The solution was defined to be
analytical outside the computational domain for simplicity. We chose $M=1$
in our scheme in Eq. (13), and the number of grid points in $x$ and $y$ to
be $200$ with $\Delta z=0.01$. Our G-FDTD methods are used to evolve Eq. (%
\ref{initconds}) in the $z$-direction simulating the propagation of the
soliton into the material after being injected at $z=0$ (see Fig. 1). In all
of our simulations, after sufficient propagation in the $z$ direction $f(z,y)
$ reaches steady-state. This solution was then substituted into Eq. (\ref%
{aequation}) to obtain the vector potential $\mathbf{A}(\mathbf{r},t)$. The
electric field within the monolayer was then calculated according to $%
\mathbf{E}(\mathbf{r},t)=-\partial \mathbf{A}(\mathbf{r},t)/\partial t$,
giving our results as shown in Fig. 2. It can be seen from Fig. 2 that our
G-FDTD method has produced the similar result as obtained by Nesterov et al.
(see Fig. 2 in \cite{Nesterov}).

In order to study the effect of point defects in a free-standing nonlinear
layer on bright optical soliton behavior, $g$ in Eq. (3) was varied such
that at $y=0,$ $g=0.5,$ and everywhere else $g=0.05.$ As before, the optical
field was injected with the profile as given in Eq. (\ref{initconds}) at $z=0
$ and was evolved in $z$ according to Eq. (\ref{schrodinger}) by the G-FDTD
method. We found that beyond $z>1$, the auxiliary function $f(z,y)$ remains
the same for all $z$, and thus steady-state has been reached. The electric
field $|\mathbf{E}(\mathbf{r},t)|$ is then calculated for the times $0\leq
t\leq 10$ in increments of $t=2$ as seen in Fig. 3. It can be seen that the
general soliton shape is preserved even in the presence of the defect, with
dips present corresponding to its location. The oscillation in time is seen
for the same reasons as Fig. 2, where the electric field oscillates due to
the energy exchange with the magnetic field. We observed no significant
scattering of the optical soliton due to the defect, with its overall shape
being unchanged during oscillation of the electric field over time. This can
be interpreted as an analogous effect to superfluidity \cite{pitaevskii03},
where a fluid can flow with reduced scattering in the presence of defects.
In a standard fluid the presence of defects creates scattering which
eventually impedes the propagation of the fluid by reflections which
randomize the momentum of the incident wave. In a superfluid below the
critical velocity, such backscattering events become energetically
suppressed leading to resistanceless flow. Similar effects in polaritonic
media have been observed \cite{wouters10}.

Furthermore, we examined the robustness of the soliton propagation in the
presence of multiple point defects placed in a free-standing monolayer
configuration. This was performed numerically in the same way by varying $g$
in Eq. (3) such that at the $y-$location of the defect, $g=0.5,$ and
everywhere else $g=0.05.$ As see in the single defect case, the bright
soliton preserves its shape despite presence of the graphene defect(s). The
electric field $|\mathbf{E}(\mathbf{r},t)|$ at $t=0$ as seen in Fig. 4
demonstrates the presence of zero to five point defects. The numerical
solutions show very stable configurations despite starting from the initial
conditions Eq. (\ref{initconds}). The steady state configuration is reached
after numerical evolution to $z=1$ and remains unchanged beyond this
distance. Again the the solitons show oscillations in the time domain, hence
we show results for $t=0$. The location of the defects create local dips in
the magnitude of the electric field but otherwise show the same soliton
profile as the defect-free solution. These results show that the graphene
monolayer serves as a highly robust material that can propagate optical
solitons even in the presence of defects. The results also show that the
explicit G-FDTD can handle strong variations in $g$ along $y$ without
diverging. Similar finite element methods tend to give diverging results for
spatially varying $g$, particularly for large variations such as considered
here.

\section{Conclusion}

We have performed a numerical study of bright optical solitons, and shown
how various material configurations can affect behavior of the electric
field along a free-standing nonlinear layer, including defects. It was found
that optical solitons can travel in such monolayers taking advantage of the
natural nonlinearity that is present in these materials. Our simulations
indicate that even in the presence of defects, the solitons can travel large
distances with the overall shape of the solitons being unchanged beyond
local variations at the defect position. This can be interpreted to be
caused by the beneficial effects of the nonlinearity. The nonlinear
interaction gives a suppressed scattering due to an effect analogous to
superfluidity. This allows the optical soliton to propagate even in the
presence of the defects, which would otherwise destroy the soliton due to
multiple backscattering events. These results show that beyond the
self-guiding nature of the nonlinear monolayer, such materials are
beneficial as they are rather robust even in the presence of graphene
imperfections.

The numerical simulations were performed using the recently developed G-FDTD
method for solving an analogue of the non-homogeneous vector Helmholtz's
equation. The large spatial variations in $g$ are efficiently simulative
thanks to the improved stability of the G-FDTD algorithm. Comparable
simulations using standard finite element methods would require much smaller
evolution steps $\Delta z$ in order to obtain convergent results. Results
show that the G-FDTD scheme provides an accessible technique for studying
dynamic solutions of the magnetic vector potential $\mathbf{A}(\mathbf{r},t)$
and electric field $\mathbf{E}(\mathbf{r},t)$ corresponding to a
free-standing graphene monolayer configuration in $3+1$ dimensions. This
makes the method appropriate suitable for studying more complex geometrical
configurations such as optical networks based on the monolayers. We leave
this and other extensions of the study as future work.

\section*{\hspace{-0.23in}Acknowledgments}

This research is supported by grant \#FA8650-05-D-1912 from the United
States Air Force Office of Scientific Research and a grant from the NASA
EPSCoR \& LaSPACE, Louisiana. The second author is supported by the Okawa
Foundation.

\begin{figure}[h!]
\centering
	\includegraphics[width=\columnwidth]{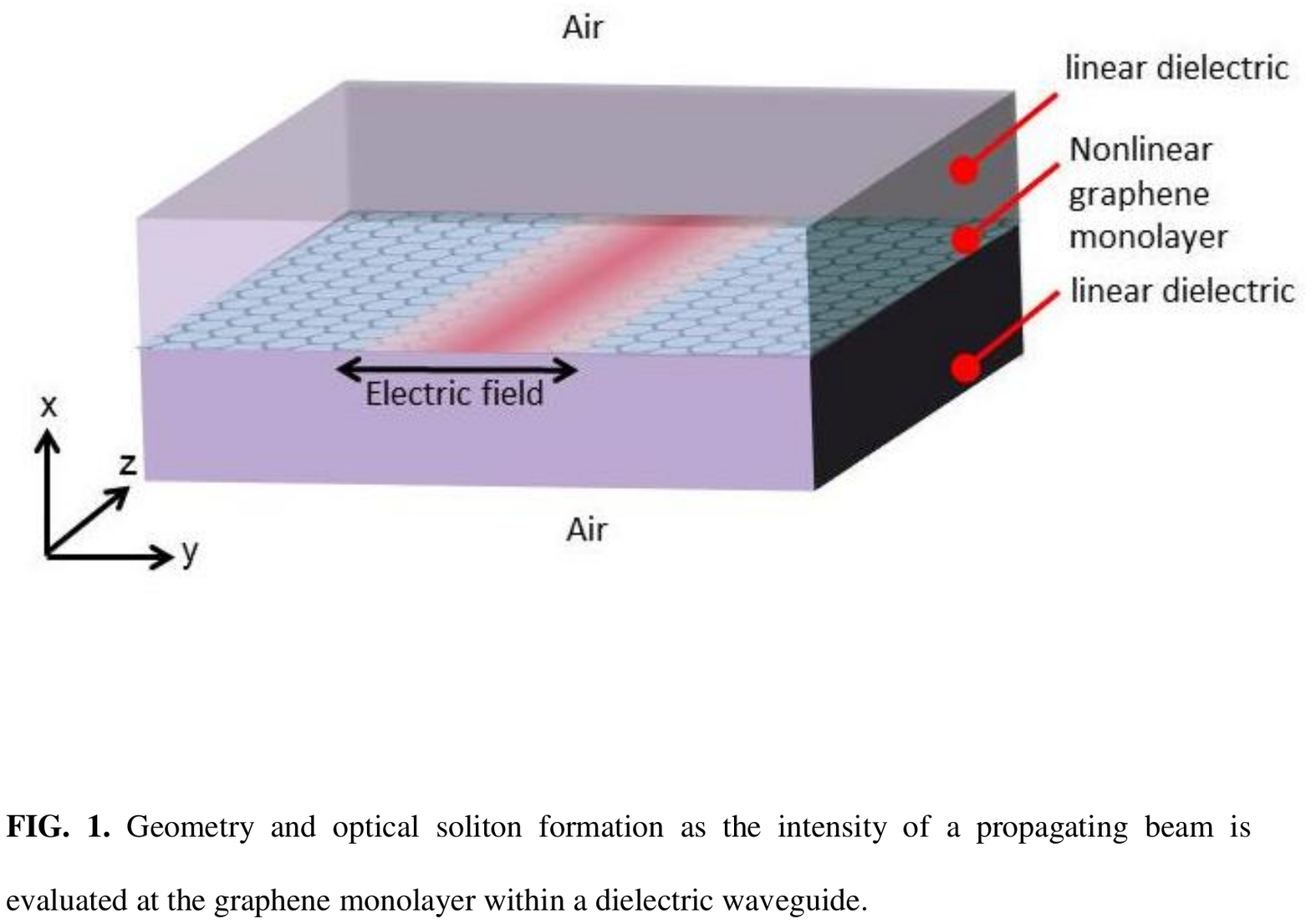}
	\label{Fig1}
\end{figure}

\begin{figure}[h!]
\centering
	\includegraphics[width=\columnwidth]{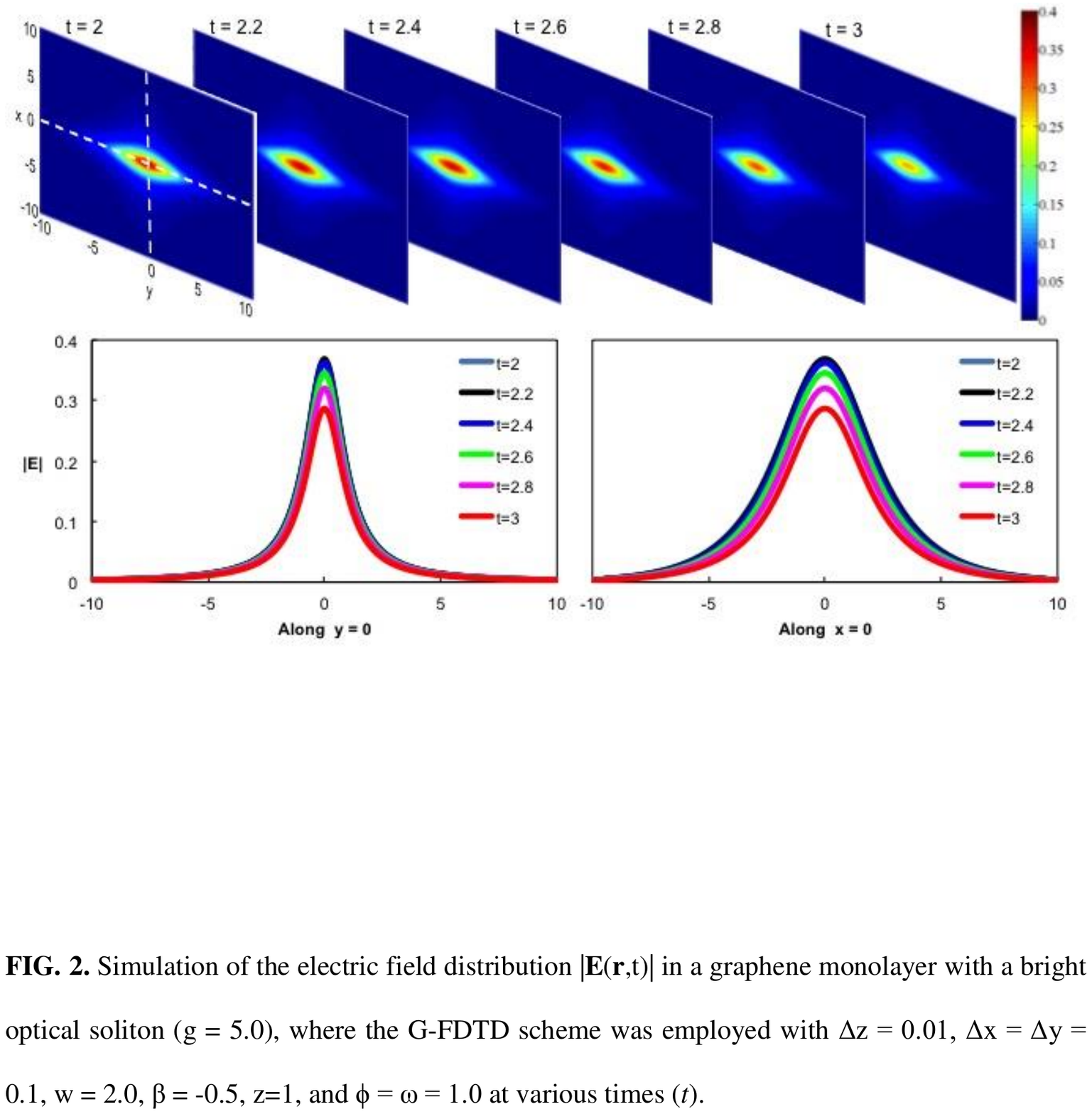}
	\label{Fig1}
\end{figure}

\begin{figure}[h!]
\centering
	\includegraphics[width=\columnwidth]{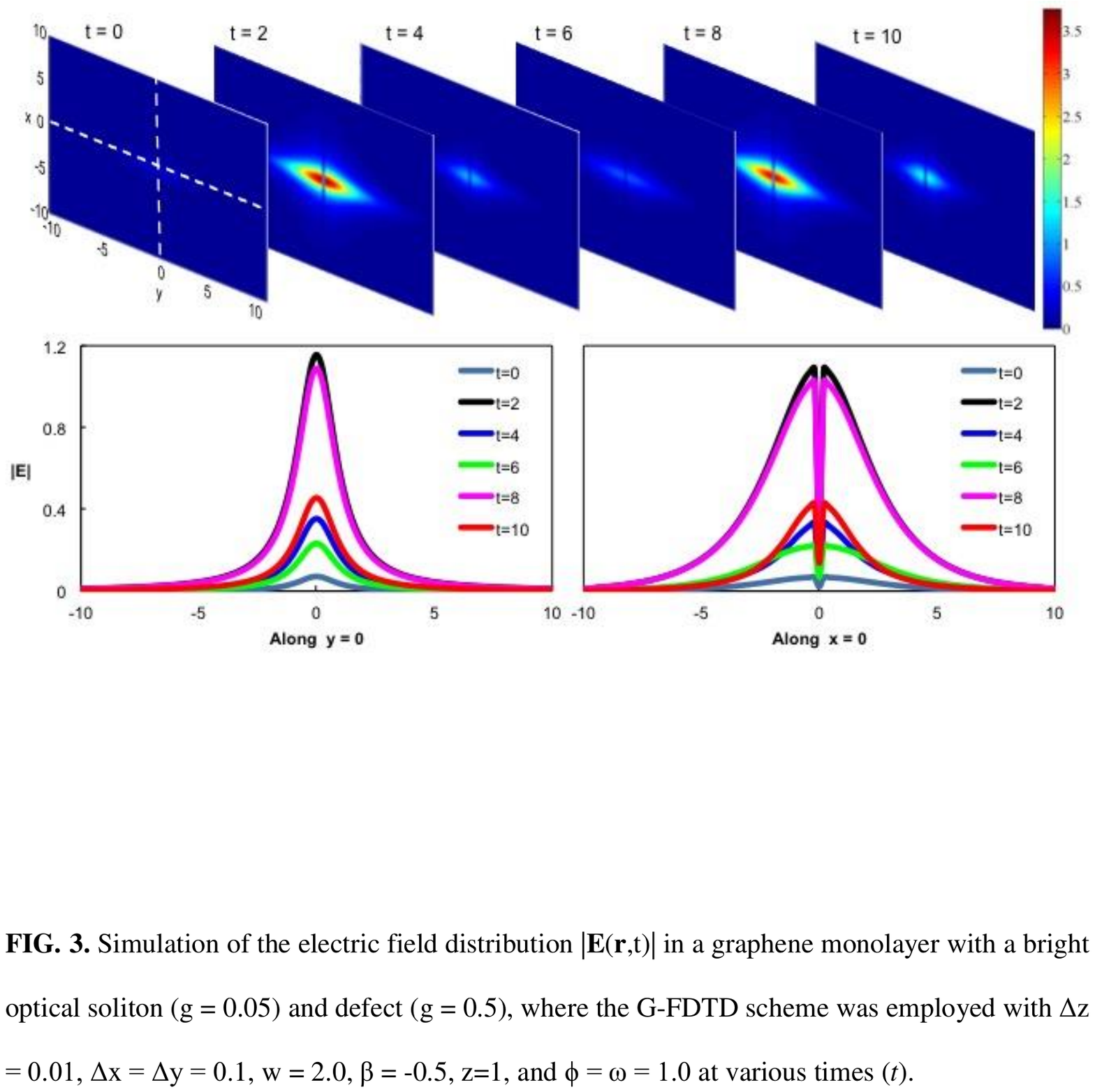}
	\label{Fig1}
\end{figure}

\begin{figure}[h!]
\centering
	\includegraphics[width=\columnwidth]{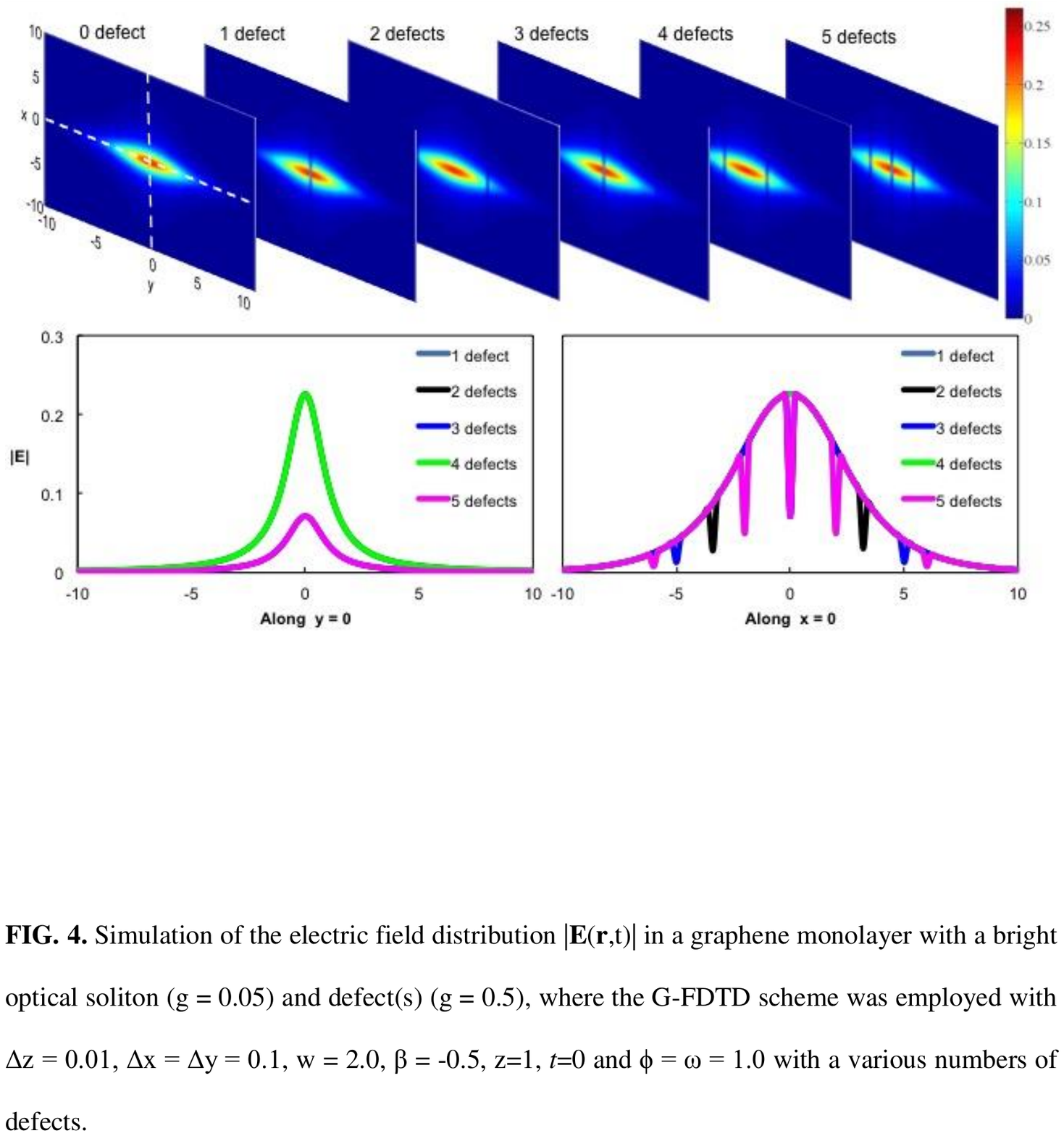}
	\label{Fig1}
\end{figure}

\end{document}